\documentclass[prl,twocolumn,superscriptaddress,showpacs]{revtex4}
\usepackage{amsmath,amssymb,graphicx}

\def\ketpsi{\vert \psi \rangle}
\def\brapsi{\langle \psi \vert}
\def\ketpsi0{\vert \psi_0 \rangle}
\def\brapsi0{\langle \psi_0 \vert}

\input{tcilatex}
\begin{document}

\title{Energy transport in strongly disordered superconductors and magnets}
\author{L. B. Ioffe}
\affiliation{Center for Materials Theory, Department of Physics and Astronomy, Rutgers
University 136 Frelinghuysen Rd, Piscataway NJ 08854 USA}
\affiliation{CNRS and Universit\'e Paris-Sud, UMR 8626, LPTMS, Orsay Cedex, F-91405 FRANCE}
\author{Marc M\'ezard}
\affiliation{CNRS and Universit\'e Paris-Sud, UMR 8626, LPTMS, Orsay Cedex, F-91405 FRANCE}
\date{\today}

\begin{abstract}
We develop an analytical theory for quantum phase transitions driven by
disorder in magnets and superconductors. We study these transitions with a
cavity approximation which becomes exact on a Bethe lattice with large
branching number. We find two different disordered phases, characterized by
very different relaxation rates, which both exhibit strong inhomogeneities
typical of glassy physics.
\end{abstract}

\maketitle

The zero temperature quantum phase transitions and their quantum critical
points have been rather well understood in translationally invariant systems%
\cite{Sachdev2000}. Much less is known about disordered systems where the
transition is driven by the competition between strong disorder and
interactions.

Motivated by experiments on disordered superconductors \cite{fn00}, we
formulate a theoretical model of disorder driven transitions and solve it in
the simplest controlled approximation. Our main results can also be relevant
for other disordered quantum problems, especially for strongly disordered
magnets. The new physics introduced by the strong disorder is the appearance
of new phases \cite{Basko2006} in which all or some excitations are
localized in space and have infinite lifetime and thus cannot contribute to
any transport. The quantum critical point at which the long range order
appears has many features that distinguish it from conventional quantum
critical points in translationaly invariant systems. Most notably, it is
characterized by a wide distribution of the order parameter and the
appearance of a new intermediate phase in which only low energy local
excitations have infinitely long lifetime while high energy ones can decay.

Strongly disordered films of InO, TiN or Be display a transition from the
superconductor to insulator when their resistivity in the normal state
exceeds a value of the order of resistance quantum $R_{Q}=6.5k\Omega$\cite%
{SITReview}. Close to the transition the superconductor-insulator phase can
also be induced by magnetic field; this transition displays a quantum
critical point behavior \cite{Kapitulnik2008}. In the vicinity of the
quantum critical point, the tunneling spectroscopy shows a well defined gap
at all points. However, the coherence peaks expected for a BCS
superconductor appear at some locations and disappear at others, a
phenomenon which is similar to some experiments in high $T_{C}$ oxides. The
absence of a coherence peak combined with the detection of a superconducting
gap in a single electron tunneling experiment implies that the disorder does
not affect the local Cooper pairing of electrons but prevents the formation
of a coherent state of these pairs. This allows to exclude `fermionic'
mechanisms of the superconductivity suppression through a reduction of
phonon attraction by Coulomb interactions. Because Coulomb interaction is
strongly suppressed in the insulating phase\cite{fn0}, the most plausible
mechanism for the superconductor-insulator transition in homogeneous
disordered films is a competition between pair hopping and random pair
energies on different sites, as suggested in a seminal paper of Ma and Lee%
\cite{MaLee1985}.

As we show in this paper the solution of this model, which requires going
beyond the simple mean-field anlaysis of the earlier works, reproduces
correctly the most important features of the data on disordered films:
direct superconductor-insulator transition, activated behavior close to the
quantum critical point in the insulating phase, strong dependence of the
activation energy near the quantum critical point and huge order parameter
variations from site to site in the superconducting phase.

In the absence of Coulomb repulsion the electrons are paired even on
localized single electron states, and pairs can hop from one site to
another. This physics is described by a Hamiltonian of disordered bosons
with strong on-site repulsion \cite{Anderson1959,MaLee1985,Feigelman2009}: 
\begin{equation}
H=-\left( \sum_{i}\xi _{i}\sigma _{i}^{z}+\sum_{(ij)}M_{ij}(\sigma
_{i}^{+}\sigma _{j}^{-}+\sigma _{i}^{-}\sigma _{j}^{+})\right)  \label{H_A}
\end{equation}

Here the state with $\sigma _{i}^{z}=\pm 1$ corresponds to a local level
occupied or unoccupied by a Cooper pair; $\xi _{j}$'s are occupation
energies for each site, which are quenched random variables drawn from a
probability $P(\xi )$. $M_{ij}$ describe the pair hopping amplitude between
sites $i$ and $j$. These hopping amplitudes couple a typical local level to
a large number of neighbors, $Z\gg 1$. We shall assume that each site is
coupled to $Z$ neighbours with $M_{ij}=g/(Z-1)$, and for technical
simplicity we shall study the slightly simplified Hamiltonian 
\begin{equation}
H=-\left( \sum_{i}\xi _{i}\sigma _{i}^{z}+\frac{g}{Z-1}\sum_{(ij)}\sigma
_{i}^{x}\sigma _{j}^{x}\right)  \label{H_B}
\end{equation}%
but all our conclusions also hold for the case (\ref{H_A}). With a
redefinition of the meaning of occupied and empty, one can take the site
energies $\xi _{j}$ to be all positive. We shall assume that the $\xi $ are
uniformly distributed on the interval $[0,1/\nu ]$. The important feature of
this distribution is that it is constant near to $\xi =0$; the value of $\nu 
$ just sets the scale of energies, and we can choose $\nu =1$. In this
language of Hamiltonians (\ref{H_A},\ref{H_B}) the superconducting phase is
mapped onto the phase with spontaneous magnetization in the $x$ direction,
in the insulating phase spins point parallel to z-axis.

The most obvious approach to study this Hamiltonian is through a simple
mean-field (SMF) approach, where $H$ is replaced by $H_{MF}=\sum_{i}(-\xi
_{i}\sigma _{i}^{z}-B\sigma _{i}^{x} ) $ and $B$ is determined
self-consistently. At temperature $T=1/\beta $, this predicts a phase
transition from insulator to superconductor at the critical value of the
hopping 
\begin{equation}
g_{c}^{{SMF}}=\left(\int d\xi P(\xi )\tanh (\beta \xi )/\xi\right )^{-1}\ .
\label{gc_naive}
\end{equation}
As $P(0)>0$, $g_{c}^{{SMF}}\to 0$ at low temperatures.

While these SMF predictions are correct at $Z=\infty $, they are
qualitatively wrong at low temperature in finite connectivity systems. We
now turn to a more refined mean field discussion, valid for finite $Z\gg 1$,
which is the basis for our main results. We use a quantum version of the
cavity method \cite{MezPar} which would become exact if the spins were on a
Bethe lattice of connectivity $Z$. In this method, one studies the
properties of a spin $j$ in the cavity graph where one of its neighbours has
been deleted, assuming that the $K=Z-1$ remaining neihbours are
uncorrelated. The system of spin $j$ and its $K$ neighbours is thus
described by the local Hamiltonian 
\begin{equation}
H_{j}^{cav}=-\xi _{j}\sigma _{j}^{z}-\sum_{k=1}^{Z-1}\left( \xi _{k}\sigma
_{k}^{z}+B_{k}\sigma _{k}^{x}+\frac{g}{K}\sigma _{j}^{z}\sigma
_{k}^{z}\right)  \label{Hcav}
\end{equation}%
where $B_{k}$ is the local \textquotedblleft cavity\textquotedblright\ field
on spin $k$ due to the rest of the spins (in absence of $j$). By solving the
problem of $Z$ Ising spins in (\ref{Hcav}), one can compute the induced
magnetization of $j$, $\langle \sigma _{j}^{x}\rangle $, which is by
definition equal to $B_{j}/\sqrt{\xi _{j}^{2}+B_{j}^{2}}$. We thus get a
mapping allowing to compute the new cavity field $B_{j}$ in terms of the $K$
fields $B_{k}$ on the neighbouring spins. This mapping induces a
self-consistent equation for the distribution of the $B$ fields \cite{MezPar}%
. We have made one more approximation which is to study the cavity
Hamiltonian (\ref{Hcav}) with a mean field method\cite{fn2}. This gives the
explicit mapping: 
\begin{equation}
B_{i}=\frac{g}{K}\sum_{k=1}^{K}\frac{B_{k}}{\sqrt{B_{k}^{2}+\xi _{k}^{2}}}%
\tanh \beta \sqrt{B_{k}^{2}+\xi _{k}^{2}}\ .  \label{eq:mapping_Kfinite}
\end{equation}%
In order to understand this mapping, let us imagine that we iterate it $R$
times on a Bethe lattice. For $R$ finite, when the number of spins is large,
the corresponding graph is a rooted tree with branching factor $Z-1$ at each
node and depth $R$. The field $B_{0}$ at the root is a function of the $%
K^{R} $ fields on the boundary. In order to see whether there is spontaneous
ordering, we study the value of $B_{0}$ in linear response to infinitesimal
fields $B_{i}=B\ll 1$ on the boundary spins. This is given by 
\begin{equation}
B_{0}/B=\Xi \equiv \sum_{P}\prod_{n\in P}\left[ \frac{g}{K}\frac{\tanh
(\beta \xi _{n})}{\xi _{n}}\right] \ .  \label{Xi_def}
\end{equation}%
where the sum is over all paths going from the root to the boundary, and the
product $\prod_{n\in P}$ is over all sites along the path $P$. The response $%
\Xi $ is nothing but the partition function for a directed polymer (DP) on a
tree, where the energy of each site is $e^{-E_{n}}=(g/K)(\tanh (\beta \xi
_{n})/\xi _{n})$ and the temperature has been set equal to one. The solution
of this problem, found in \cite{DerrSpohn} , can be expressed in terms of
the convex function $f(x)=(1/x)\log \left[ K\int_{0}^{1}d\xi \;\left( \tanh
(\beta \xi )/\xi \right) ^{x}\right] $, which is minimal at a value $x=x_{c}$%
. In the large $R$ limit, there exist two phases for the DP problem:

\begin{itemize}
\item ``Self-averaging'' (SA) phase: If $x_c>1$, then $(1/R)\log\Xi=f(1)+%
\log(g/K) $. The ordered phase appears at $g_c=K e^{-f(1)}=g_c^{SMF}$.

\item ``Glassy phase'' (G) phase: If $x_c<1$, then $(1/R)\log\Xi=f(x_c)+%
\log(g/K) $. The ordered phase appears at $g_c=K e^{-f(x_c))}>g_c^{SMF}$.
\end{itemize}

These two regimes of the DP problem are qualitatively very different. The
\textquotedblleft SA\textquotedblright\ regime is the high \textquotedblleft
temperature\textquotedblright\ phase of the polymer, where the measure on
paths defined in (\ref{Xi_def}) is more or less evenly distributed among all
paths. The low temperature \textquotedblleft G\textquotedblright\ regime is
a glass phase where the measure condensates onto a small number of paths. An
order parameter which distinguishes between these phases is the
participation ratio $Y=\sum_{P}w_{P}^{2}$, where $w_{P}$ is the relative
weight of path $P$ in the measure (\ref{Xi_def}). In the replica formalism
the SA phase is replica symmetric, $\Xi $ is self-averaging, and $Y=0$; the
G phase is a one-step replica-symmetry-breaking (RSB) glass phase, the value
of $Y$ is finite and non self-averaging (it depends on the explicit
realization of the $\xi $'s even in the thermodynamic limit), and its
average is given by $1-x_{c}$\cite{CookDerrida90}. This glass transition,
and the nature of the G phase, are identical to the ones found in the random
energy model \cite{DerridaREM,GrossMezard}.

\begin{figure}[ht]
\includegraphics[width=7 cm,angle=0]{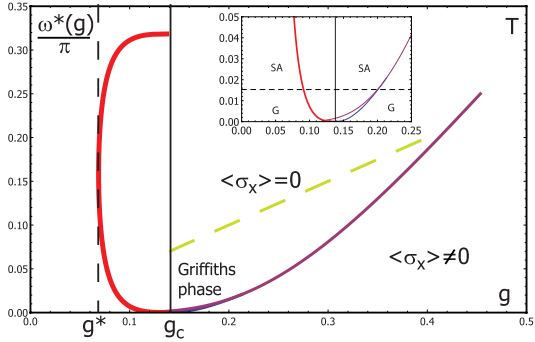}
\caption{Phase diagram of the spin system for $K=2$. The right pane shows
the critical line $T_c(g)$ while the left one shows the critical energy that
separates the states with zero width from those with a finite width. The
inset shows the low temperature/energy region.}
\label{fig:phase_diag}
\end{figure}

Using these DP results one gets the phase diagram of the spin systems shown
in the right pane of Fig.\ref{fig:phase_diag}. At any temperature, there is
a non-zero critical value of the coupling, $g_{c}(T)$, separating an
\textquotedblleft ordered\textquotedblright , superconducting phase with
spontaneous $x$ magnetization at $g>g_{c}(T)$ from a normal
\textquotedblleft disordered\textquotedblright , insulating phase with zero
magnetization at $g<g_{c}(T)$. Within each phase, there are two regimes of
temperature, \textquotedblleft Self-averaging\textquotedblright\ and
\textquotedblleft Glassy\textquotedblright . As is clear from our
susceptibility analysis, the glass transition of the DP affects the
propagation of a static perturbation in the spin system. In the disordered
SA phase, the total effect of the perturbation decreases when $R$ increases,
and propagates evenly: the average value of the susceptibility coincides
with its typical value. In the disordered G phase, the total perturbation
also decays, but it condenses on a finite number of paths. Consequently, the
susceptibility is non-self averaging, similarly to what is found in one
dimension\cite{Fisher92}. Rare paths are important in the whole
\textquotedblleft Griffiths\textquotedblright\ phase where the
susceptibility distribution has a power law tail. When $g_{c}^{SMF}<g<g_{c}$
the typical susceptibility is finite but the average susceptibility
diverges. In the ordered phase, the perturbation propagates to infinity,
again with very different patterns in the \textquotedblleft
SA\textquotedblright\ and \textquotedblleft G\textquotedblright\ phases. The
SMF gets the correct result of the SA regimes, but completely misses the
low-temperature physics of condensed correlation paths.

The RSB transition also strongly affects the scaling of the field in the
ordered phase, for $g=g_{c}(1+\epsilon )$. One can study the distribution of
fields $P(B)$ induced by the mapping (\ref{eq:mapping_Kfinite}). An
expansion of its Laplace transform shows that, in the G phase, $P(B)$ decays
at large $B$ like $P(B)\simeq C/B^{1+x_{c}}$. This distribution has a
diverging mean, dominated by rare fluctuations. A careful analysis of the
self-consistent equation for $P(B)$ then indicates that the typical scale of
the field behaves as 
\begin{equation}
B_{\text{typ}}\simeq Ae^{-g_{B}/(g-g_{c})}\ .
\end{equation}

In the disordered phase the average value of the transverse field is zero,
but its quantum fluctuations become important. Their main physical effect is
the broadening of the local levels that, in the absence of $g$, correspond
to $\sigma _{i}^{z}=\pm 1$. At $T=0$ the level broadening means that local
excitations of energy $\omega \simeq 2\xi _{i}$ decay. If we neglect the
phonons, energy conservation requires that some group of spins with the same
energy be flipped. In finite systems this is generically impossible. To
study whether energy can be transported, and ergodicity can be restored, in
infinite systems, we adopt an approach similar to the one developed above.
Namely, we consider a Bethe lattice that is very weakly coupled to the
environment at its boundary and study the effective level width of a 'root'
spin at a distance $R$ from the boundary in the $R\rightarrow \infty $
limit. Thus we add to the Hamiltonian (\ref{H_B}) the boundary term $%
H_{env}=-\sum_{j}\sigma _{j}^{x}x_{j}(t)$ where $x_{j}(t)$ are \emph{%
dynamical} fields generated by the environment, characterized by a response
function $G(\omega )$. In the leading order in $g/K$ the relaxation rate of
the root spin follows from the Fermi golden rule: 
\begin{equation}
\Gamma _{0}(\omega )=\func{Im}G(\omega )\sum_{P}\prod_{n\in P}\left[ \frac{%
2g/K}{\omega -2\xi _{k}}\right] ^{2}\ .  \label{Gamma_Linear}
\end{equation}%
This perturbation-theory-based equation is only valid when all fractions
inside the product remain small\cite{fn4}, and the relaxation rate of each
spin is very small. Thus it is self-consistent if $\Gamma _{0}\rightarrow 0$
when $R\rightarrow \infty $. The typical value of $\Gamma _{0}(\omega )$ is
controlled by 
\begin{equation*}
f_{\Gamma }(\omega )=\frac{1}{R}\overline{\ln \left\{ \sum_{P}\prod_{n\in P}%
\left[ \frac{2}{\omega -2\xi _{k}}\right] ^{2}\right\} }\ ,
\end{equation*}%
it decreases away from the boundary if $f_{\Gamma }(\omega )+2\ln (g/K)<0$. $%
f_{\Gamma }$ can be computed again by analyzing a DP problem, which turns
out to be always in its G phase. At $\omega =0$, one finds that $f_{\Gamma
}(0)+2\ln (g/K)\leq 0$ in the whole insulating regime, and the equality is
reached at the critical point $g=g_{c}$. One also finds that the region of
small $\omega -2\xi $ gives negligible contribution to $f_{\Gamma }(\omega )$%
, which allows one to work with the unregularized expression (\ref%
{Gamma_Linear}). At non-zero frequencies $f_{\Gamma }(\omega )$ decreases,
it is minimal at $\omega =1/2$ (which corresponds to the center of the band
in our notations). At $g<g^{\ast }=Ke^{f_{\Gamma }(1/2)/2}$ the relaxation
rate is zero for all states, this is the superinsulator regime of \cite%
{Basko2006}. In the intermediate regime $g_{c}(0)>g>g^{\ast }$ the states in
the middle of the band have finite width, they are separated from the
zero-width states by a critical energy $\omega ^{\ast }(g)$ similar to the
mobility edge of the non-interacting problem.

In order to understand the low temperature properties of the system, it is
important to know the dependence of $\Gamma (\omega )\ $ at $\omega >\omega
^{\ast }(g)$. In this regime (\ref{Gamma_Linear}) must be modified. Using
the mapping to fermions as in \cite{fn4}, one finds the iterative equation
for the width of the levels on the Bethe lattice: 
\begin{equation}
\Gamma _{i}(\omega)=(2g/K)^{2}\sum_{k(i)}\frac{\Gamma _{k}(\omega)}{\left(
\omega -2\xi _{k}\right) ^{2}+\Gamma _{k}(\omega)^{2}}
\label{Gamma_Nonlinear}
\end{equation}

This equation is similar to the equation (\ref{eq:mapping_Kfinite}) for the
fields in the ordered phase and can be analyzed with the same method, giving
the fast level-width dependence slightly above $\omega ^{\ast }(g)$: 
\begin{equation}
\Gamma _{\text{typ}}(\omega)\simeq {\Gamma^{\ast }} e^{-\omega
_{0}(g)/(\omega -\omega ^{\ast }(g))}\ .  \label{Gamma_typical}
\end{equation}

This behaviour has important consequencies for the low temperature
properties of the relaxation, as we now discuss. A non-zero but low
temperature affects the relaxation rate in several ways. First, it changes
the occupation numbers of the excited and ground states, this affects the
perturbative equation (\ref{Gamma_Linear}) and thus shifts the position of
the $\omega ^{\ast }(g)$ line, this effect is however small at $T\ll 1.$
More importantly, a non-zero temperature creates some mobile excitations
with frequencies above $\omega ^{\ast }(g)$. These excitations provide a
mechanism for a small level broadening of the very low frequency levels:
They see a mobile excitation with energy $E$ with an Arrhenius rate, giving
a width $\exp (-\omega _{0}/(E-\omega ^{\ast })-E/T).$ The dominant
contribution comes from energies $E=\omega ^{\ast }(g)+\sqrt{\omega _{0}T}$,
and results in the temperature dependence $\Gamma (g)\sim \exp (-2\sqrt{%
\omega _{0}/T}-\omega ^{\ast }(g,T)/T)$ that shows a crossover between a
square root and activated or even faster behavior as one goes away from the
critical point.

In conclusion, we have outlined a solution of the strongly disordered spin
model on the Bethe lattice which shows a series of two zero-temperature
transitions between a phase with no relaxation, a phase with a slow
relaxation and an ordered phase. It also shows that the low temperature
phases are always very strongly non-uniform: both the order parameter
formation and the spin relaxation are controlled by rare interaction paths
containing a very small number of spins. When applied to the
superconductor-insulator transition our results imply the existence of both
weak and strong insulators. At the critical point the relaxation rate varies
as $\exp (1/\sqrt{T})$ but crosses over to activated at lower $g$ and low $T$%
, in the strong insulator the relaxation is completely suppressed. Of
course, some of the physical effects neglected in our model would lead to a
very slow relaxation even in the strong insulator.

We acknowlege very useful discussions with M. Feigel'man and with M.
Mueller, who has obtained related results from phenomenological
considerations \cite{muller}. This research was supported by Triangle de la
physique 2007-36, ANR-06-BLAN-0218, ARO 56446-PH-QC and DARPA
HR0011-09-1-0009.


\begin{thebibliography}{99}
\bibitem{fn0} The situation is very different in Josephson arrays and
granular films which display very different phenomenology \cite{fn00}.
\bibitem{fn2} The use of the mean field approximation in the cavity mapping
is very different from its use in (\ref{H_B}): as we show elsewhere \cite%
{fn00}, it captures the leading $1/Z$ effects, even though it neglects some
time correlations and level repulsion.
\bibitem{fn3} The ``temperature'' of the DP is here equal to $1$ and has
nothing to do with the inverse temeprature $\beta$ of the spin problem.
\bibitem{fn4} When some of the energy denominators get small, they are
regularized to $(\omega -2\xi _{n})^{2}+(g/K)^{2}$. This can be shown by
noticing that at most one neighbor of a given spin has a small energy
denominator. This reduces the problem to a 1D spin chain in which individual 
$\xi $ are distributed with the probabilty density $K$ around $\omega /2$
which can be mapped onto non-interacting fermions.
\end{thebibliography}
\end{document}